# Reimagining Sense Amplifiers: Harnessing Phase Transition Materials for Current and Voltage Sensing

Md Mazharul Islam, Shamiul Alam, Mohammad Adnan Jahangir, *Graduate Student Member, IEEE,* Garrett S. Rose, Suman Datta, Vijaykrishnan Narayanan, Sumeet Kumar Gupta, Ahmedullah Aziz, *Senior Member, IEEE*

*Abstract*— Energy-efficient sense amplifier (SA) circuits are essential for reliable detection of stored memory states in emerging memory systems. In this work, we present four novel sense amplifier (SA) topologies based on phase transition material (PTM) tailored for non-volatile memory applications. We utilize the abrupt switching and volatile hysteretic characteristics of PTMs which enables efficient and fast sensing operation in our proposed SA topologies. We provide comprehensive details of their functionality and assess how process variations impact their performance metrics. Our proposed sense amplifier topologies manifest notable performance enhancement. We achieve a ~67% reduction in sensing delay and a ~80% decrease in sensing power for current sensing. For voltage sensing, we achieve a ~75% reduction in sensing delay and a ~33% decrease in sensing power. Moreover, the proposed SA topologies exhibit improved variation robustness compared to conventional SAs. We also scrutinize the dependence of transistor mirroring window and PTM transition voltages on several device parameters to determine the optimum operating conditions and stance of tunability for each of the proposed SA topologies.

## I. INTRODUCTION

Memories are one of the most fundamental components in modern computing architecture, occupying a significant portion of the chip's area and accounting for a significant share of the overall power consumption [1], [2]. Hence, the performance of the overall system is heavily dependent on the memory which has prompted an intensive focus on the optimization of the memory array [3]. Additionally, there has been a concentrated exploration of innovative memories in alternative technological paradigms [4]–[10]. As part of this endeavor, there has been a surge of interest in non-volatile memory (NVM) technologies in recent years [3], [11]–[16]. However, these emerging memory technologies exhibit significant limitations in their performance due to individual device and architecture-level challenges [17]. Aside from the memory array, the creation of low-power, robust peripheral circuits to meet the memory array requirement is also crucial. Among these peripherals, the sense amplifier (SA) plays the most critical role as the reliability of the data retrieval directly depends on it [18]–[20]. The varying operating voltage range and sensing mechanisms of the emerging memory technologies pose significant intricacies in SA designing. These encompass proper biasing, achieving adequate voltage gain, and coping with data sensing challenges, etc. Similarly, certain memory technologies and architectures are better suited for current sensing rather than voltage sensing. Clearly, optimizing the performance of sense amplifiers is a matter of utmost importance to align with the demands of these emerging memory technologies.

Numerous research efforts have been explored to improve the efficiency, latency, and robustness of the sensing operation [21]–[25]. Some approaches employ circuit-level modifications to counter the associated challenges [23]. Conversely, others leverage distinct memory element features to implement novel sensing schemes, aiming to ease the design constraints imposed on sense amplifier circuits [24]. Moreover, the generation of more reliable reference voltages and currents has also been explored [25]. Each of these techniques has its own set of trade-offs. Thus, there is a pressing need for innovative methodologies to develop low-power, resilient sense amplifiers compatible with new and emerging memory technologies.

Prior research predominantly employs circuit-based techniques to refine sense amplifier designs. One such notable effort is the sense amplifier design based on phase transition materials (PTM). PTM has garnered significant interest from device researchers due to its several unique advantages such as abrupt threshold switching, hysteresis, high selectivity, and CMOS compatibility [4], [26]. For these advantages, PTM has been used with a conventional FET by simple augmentation to achieve steep (<60 mV) switching characteristics without any additional area penalty [27]. This structure has been commonly referred to as Hyper-FET in literature. The unique attributes of PTM and Hyper-FET were previously leveraged to design a low-power current sense amplifier [28]. In this design, a p-type Hyper-FET was emplosyed to achieve a sharp sensing mechanism. The proposed SA topology exhibits low power consumption and reduced sensing delay owing to the steep switching characteristics of the PTM. However, certain memory technologies are more suitable for voltage sensing rather than current sensing[25], [29]. Hence, it is of paramount interest to explore PTM-based voltage SA topology. Besides, it is also crucial to explore the performance of similar SA topologies based on n-type Hyper-FET to investigate the unique advantages that they can offer. In this manuscript, we propose three additional SA topologies based on PTM-augmented Hyper-FET.

In a current sense amplifier circuit, a diode-connected transistor is used for current-to-voltage conversion. A sufficient margin of voltage level corresponding to two different current levels is required for reliable sensing operation. So, it is imperative to analyze the sensing window for both diode-

Manuscript received XXXXXXXXX; accepted XXXXXXXXX. Date of publication XXXXXXXXX; date of current version XXXXXXXXXX. This material is based in part on research sponsored by Air Force Research Laboratory under agreement FA8750-21-1-1018. This work was also supported in part by NSF. Award No: 2052780.

M. M. Islam, M. A. Jahangir, Shamiul Alam, Garrett S. Rose, and A. Aziz are with the Department of EECS, University of Tennessee, Knoxville, TN, USA. (E-mail: mislam49@vols.utk.edu, salam10@vols.utk.edu, garose@utk.edu, aziz@utk.edu).
S. Datta is with the School of Electrical and Computer Engineering, Georgia Institute of Technology, Atlanta, GA, USA. (E-mail: sdatta68@gatech.edu)
V. Narayanan is with the School of Electrical and Computer Science & Engineering and Electrical Engineering, Pennsylvania State University, State College, PA, USA. (E-mail: vijay@cse.psu.edu)
S. Gupta is with the School of Electrical and Computer Engineering, Purdue University, West Lafayette, IN, USA. (E-mail: guptask@purdue.edu)

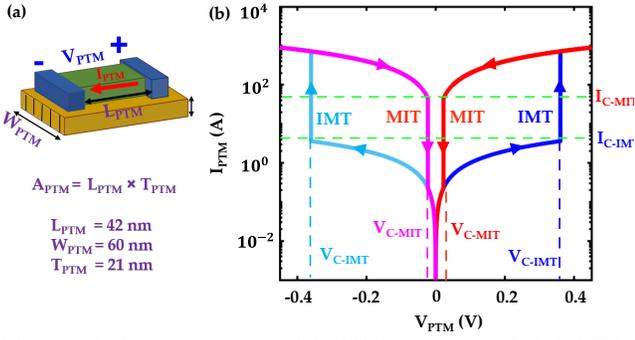

**Figure 1: (a)** Device structure, and **(b)** *I-V* characteristics of a bulk PTM.

connected PMOS and diode- connected NMOS. Again, process yield is of major concern since reference precision is a critical issue in sense amplifier circuit performance. Besides, in the current sense amplifier where current-to-voltage conversion is required, transistor mismatch may lead to inaccurate current-to-voltage conversion, thus diminishing sensing accuracy. Thus, a comprehensive variation analysis is required for the assessment of their variation robustness. The key contributions of this paper are as follows-

  1) We provide the comprehensive functionality analysis of the previously proposed current sense amplifier circuit [28].

  2) We then propose three additional SA topologies based on PTM and discuss their detailed functionality.

  3) We perform a comprehensive analysis to inspect the dependence of transistor mirroring window and PTM transition voltages on several device parameters to determine the optimum operating conditions.

  4) Finally, we perform 1000-point $3\sigma$-Monte-Carlo simulations to determine the effects of external variations.

The organization of this paper is as follows: Section II provides a brief introduction to PTM, PTM augmented Hyper-FET, and their unique characteristics. Section III discusses the construction and functioning of conventional sense amplifiers. Section IV presents the detailed functionality of proposed SA topologies. In Section V, we discuss the window analysis and transition voltage analysis of Hyper-FET with the variation of threshold voltage and the number of fins. Finally, in section VI, we present variation analysis by illustrating the Monte-Carlo simulation result and provide a comparative analysis between our proposed SA topologies and conventional SA topologies.

## II. PTM AND HYPER-FET

PTM refers to a group of materials which demonstrate abrupt change resistances due to insulator-to-metal or metal-to-insulator transition [27], [30], [31]. These transitions can be prompted by electrical, thermal, mechanical, or optical stimuli. Several physical mechanisms are responsible for the transition observed in PTM such as ion diffusion [32],[33], dimerization [34] and electron correlation [35]. Here, we mainly focus on PTM that is triggered by electrical stimuli. At a low voltage, PTM remains in high resistance or insulating state. When a sufficiently high voltage ($V_{C-IMT}$) is applied across it, insulator-to-metal transition (IMT) occurs (Figs. 1(a,b)) [36]–[38]. The resistivity in insulating state ($\rho_{INS}$) is orders of magnitude higher ssthan the resistivity in metallic state ($\rho_M$). Conversely, when the voltage across it is reduced below a sufficiently low voltage ($V_{C-MIT}$), the PTM undergoes metal-to-insulator transition. So far, different PTMs have been discovered with diverse range of transition voltages and hysteresis behavior [4].

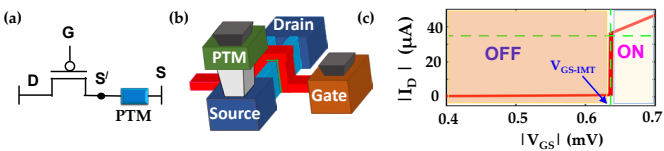

**Figure 2: (a)** Schematics, **(b)** 3-D structure, and **(c)** *I-V* characteristics of a Hyper-FET.

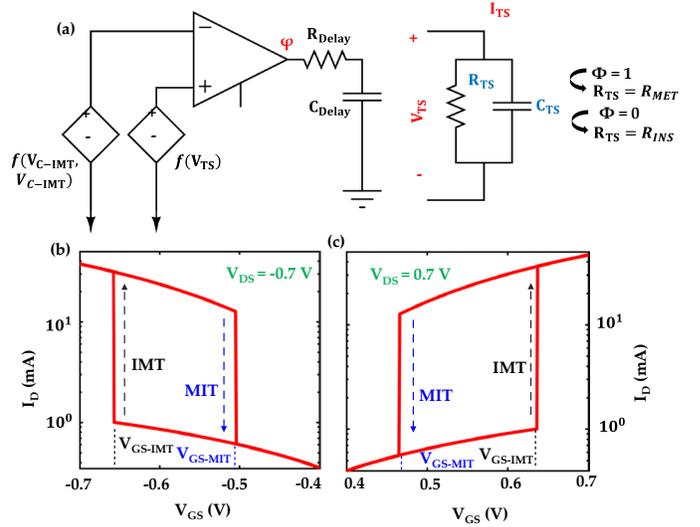

**Figure 3: (a)** Behavioral representation of the transitions observed in a PTM. $I_D$-$V_{GS}$ characteristics of **(b)** Hyper-PMOS and **(c)** Hyper-NMOS.

Hyper-FET is constructed by augmenting a PTM to the source terminal of a conventional Field effect transistor (FET) (Figs. 2(a,b)) [39]. The electrical characteristic of a Hyper-FET reflects both the abruptness of a PTM and the behavior of a conventional FET (Fig. 2(c)). When the gate-to-source voltage ($V_{GS}$) is lower than $V_{GS-IMT}$, the PTM is at the insulating state (HRS) acting as a very high resistance in source terminal and drastically reducing the off-current ($I_{OFF}$). However, when $V_{GS}$ is higher than $V_{GS-IMT}$, the PTM undergoes insulator-to-metal transition (IMT). At the metallic state, it acts as a low resistance at source terminal, keeping the intrinsic on-current ($I_{ON}$) of the transistor unaffected. This way, the Hyper-FET provides abrupt switching with high selectivity (as high as $10^8$).

In this work, we leverage the abrupt switching capability and high selectivity of the Hyper-FET to achieve efficient current and voltage sensing. For electrical modeling of the Hyper-FET, we used predictive technology model [28] to simulate 14nm Fin-FET as the host transistor. We incorporated the behavioral circuit model for vanadium-dioxide ($VO_2$) as the PTM (Fig.3(a)) [40]. The I-V characteristics of a pmos-based Hyper-FET (Hyper-PMOS) and a nmos-based Hyper-FET (Hyper-NMOS) are shown in Fig. 3(b) and Fig. 3(c), respectively.

## III. CONVENTIONAL SENSE AMPLIFIERS

Sense amplifiers can be broadly categorized into two major types: (i) voltage, and (ii) current SA. In a conventional voltage sense amplifiers (VSA) (Fig. 4(a)) [41], specific bit-line (BL) is pre-charged to a target value ($V_{PRE}$) at the beginning of the sensing cycle. After the pre-charge phase, when the word-line (WL) is made high, the BL discharges or remains at $V_{PRE}$ based on the memory state of the cell. During the sensing phase, if the cell is in logic '0' state, the BL is discharged by the cell current which corresponds to the low-resistance state (LRS) current (Fig.4(b-e)). But if the stored memory is at '1', the BL is preserved at the pre-charged value by the cell current that represents the high-resistance state (HRS) (Fig.4(f-i)). A voltage comparator then compares the voltage of the sense-node with the reference voltage ($V_{REF}$) and eventually generates a digital output. Fig. 4(a) shows the schematic of the conventional VSA and Figs. 4(b-i) illustrate in the pre-charging and sensing phases for both LRS and HRS states of the bit cell. Conventional VSA suffers from BL offset, owing to the BL load mismatch and BL noise (due to the coupling between WL and BL, crosstalk, etc). Also, both the comparator circuit and the core sensing circuit have input offsets resulting from the transistor mismatch (such as mismatch in $V_{TH}$, width and the oxide thickness) [42],[43]. Therefore, conventional VSAs require sufficient voltage swing in the BL to ensure reliable read operation. This results in a longer transitioning time for the BL voltage. Therefore, conventional VSAs suffer from sensing latency specially for the memories with low cell current and high BL load.

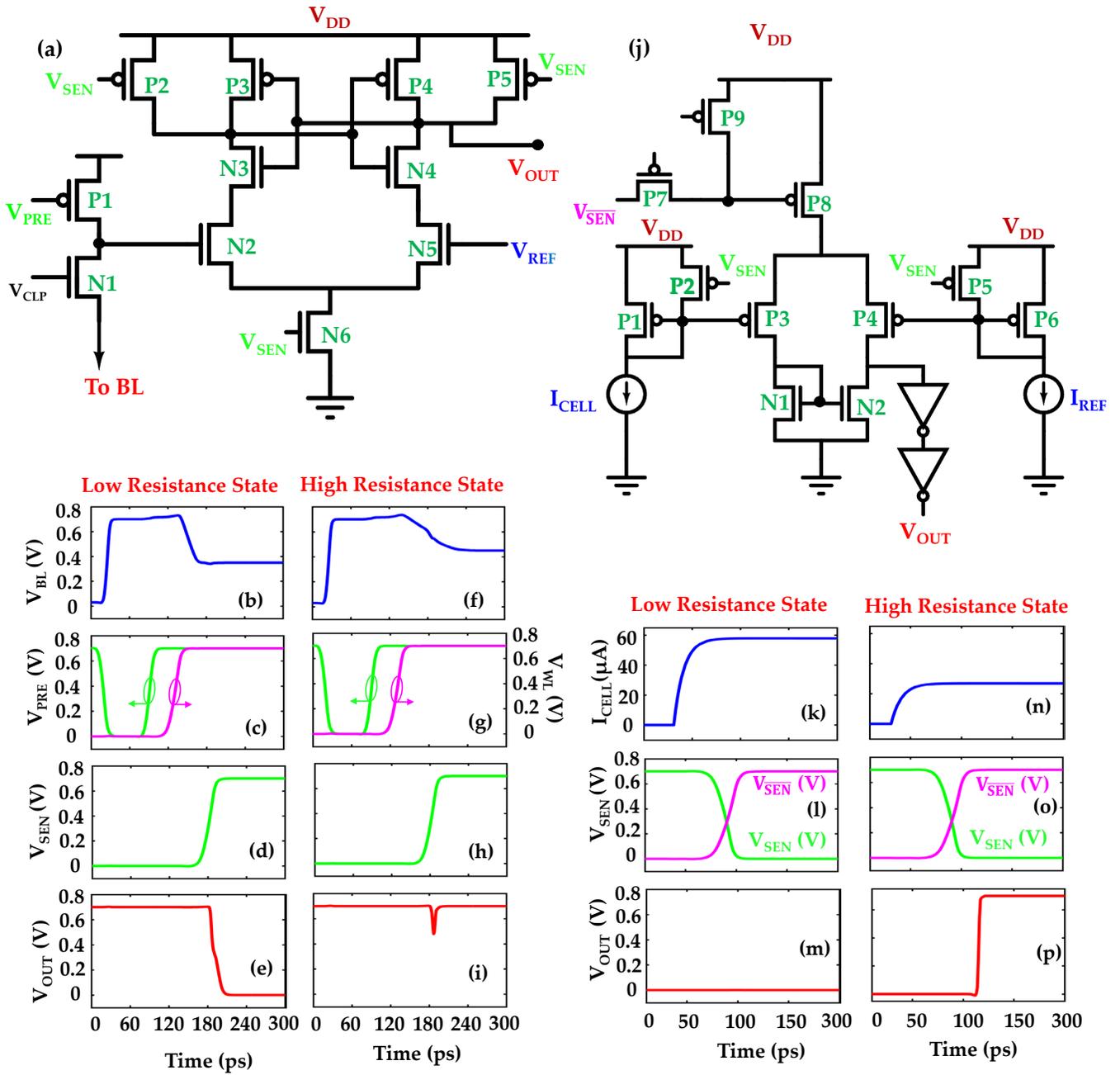

**Figure 4:** (a) Schematic of a conventional current sense amplifier. Simulation waveforms for conventional VSA during sensing (b)-(e) LRS and (f)-(i) HRS of a memory cell. (j) Schematic of a conventional current sense amplifier. Simulation waveforms for conventional CSA during sensing (k)-(m) LRS and (n)-(p) HRS of a memory cell.

On the other hand, current sense amplifiers (CSA) are mostly used for the memories with low cell current and high BL load requiring high sensing speed. Cascode-Current-Load CSA (CCL-CSA), Global-Clamping-Local-Discharging CSA (GCLD-CSA) and Current-Load CSA (CCL-CSA), Global-Clamping-Local-Discharging CSA (GCLD-CSA) and current mirror-based CSA (CM-CSA) are few of the commonly used CSA topologies [25][24][44]. Current mirror type SA (CM-CSA) based on differential input is widely used as high speed sense amplifier [25]. Fig. 4(j) shows the schematic of a conventional current mirror-based CSA [25]. The basic component of this topology is the current mirrors which make copies of the cell current ($I_{CELL}$) and the reference current ($I_{REF}$). These copied versions of $I_{CELL}$ and $I_{REF}$ are then fed to a differential amplifier which performs the I-V conversion and comparison. Based on the difference between $I_{CELL}$ and $I_{REF}$, different voltage levels are generated and further processed by the buffer stage to generate the digital output. Figs. 4(k-m) and Figs. 4(n-p) illustrate the sensing phases of conventional CM-CSA for sensing the LRS and HRS states of a memory cell, respectively. This design requires accurate transistor biasing and could be vulnerable to imbalances between the two amplifier branches. The source currents ($I_{CELL}$ and $I_{REF}$) and the mirrored currents may have an offset due to the transistor mismatches in the current mirror circuits resulting in a voltage offset between the two inputs [25]. Now, for small $I_{CELL}$ and $I_{REF}$, these offsets will result in a low sensing yield. Therefore, this topology requires accurate biasing and proper sizing of the transistors for reliable sensing operation. Considering these drawbacks, it is of great importance to design efficient sense amplifier circuits for reliable voltage and current sensing for NVM memory application.

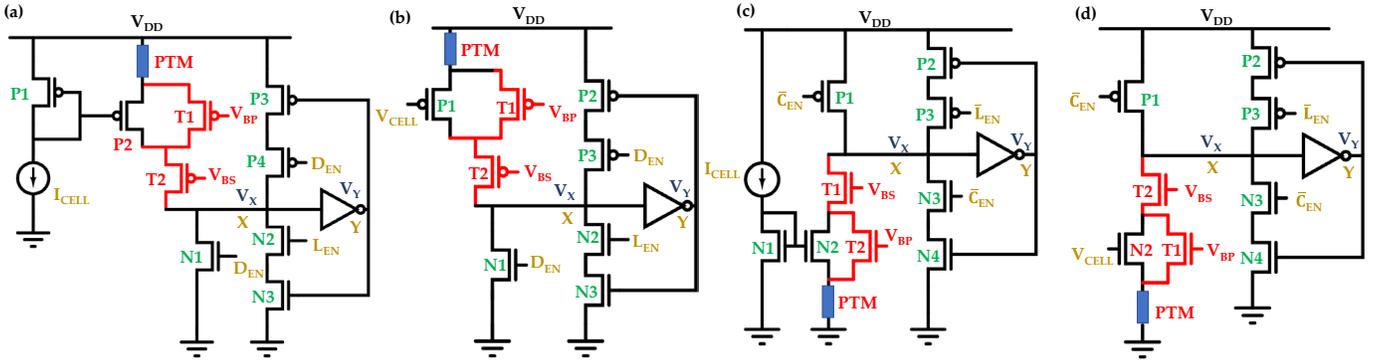

**Figure 5**: Proposed four different PTM-augmented SA topologies. Schematic of **(a)** Hyper-PMOS CSA, **(b)** Hyper-PMOS VSA, **(c)** Hyper-NMOS CSA, and **(d)** Hyper-NMOS VSA.

## IV. PROPOSED SENSE AMPLIFIER TOPOLOGIES

The previous work [28] extensively covered the PTM-based CSA (Hyper-PMOS CSA) topology, providing a comprehensive explanation of its detailed functionality. While the circuit configuration of a Hyper-NMOS VSA was introduced in the same study, its operational details were not thoroughly explained. In this section, we introduce two additional SA Circuit topologies. Each of these topologies will be thoroughly discussed and examined. Fig. 5 shows the circuit schematics of the proposed PTM-based SA topologies. Here the transistors T1 and T2 are used for tuning the reference voltage or current by controlling the voltage drop across the host transistor of the Hyper-FET. T1 is positioned in parallel with the main transistor to effectively reduce the resistivity, consequently minimizing the voltage drop across it. Thus, it increases the voltage across the PTM. In this case, IMT occurs faster but at the cost of higher power consumption. T2 is used in series which can decrease the drop across PTM. Thus, IMT will be delayed but there will be less power consumption. Hence, the incorporation of these two transistors offers the flexibility to choose either low-power/high-delay or high-power/low-delay operations while keeping the core functionality unaffected. For the sake of simplicity, we will exclude T1 and T2 in our discussion. The circuit operation of each of the topologies will be discussed in the subsequent section.

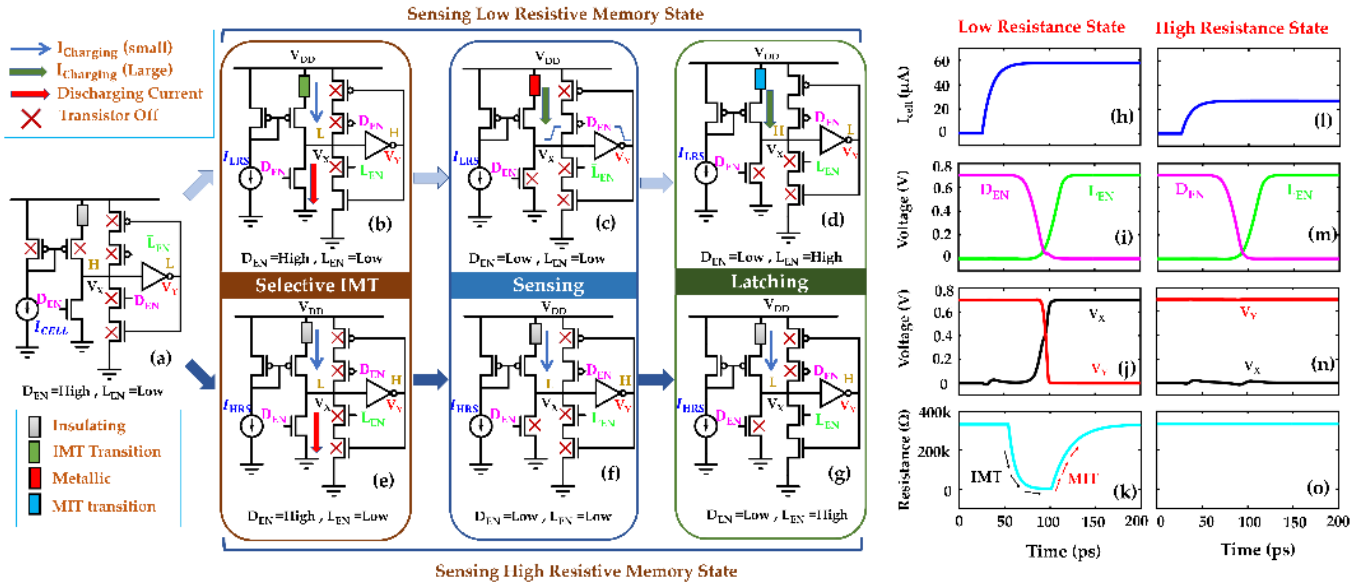

**Figure 6:** Steps of Sensing **(b)**-**(d)** LRS and **(e)**-**(g)** HRS for a Hyper-PMOS Current Sense Amplifier. Waveforms for sensing **(h)**–**(k)** LRS and **(l)**-**(o)** HRS for a Hyper-PMOS CSA

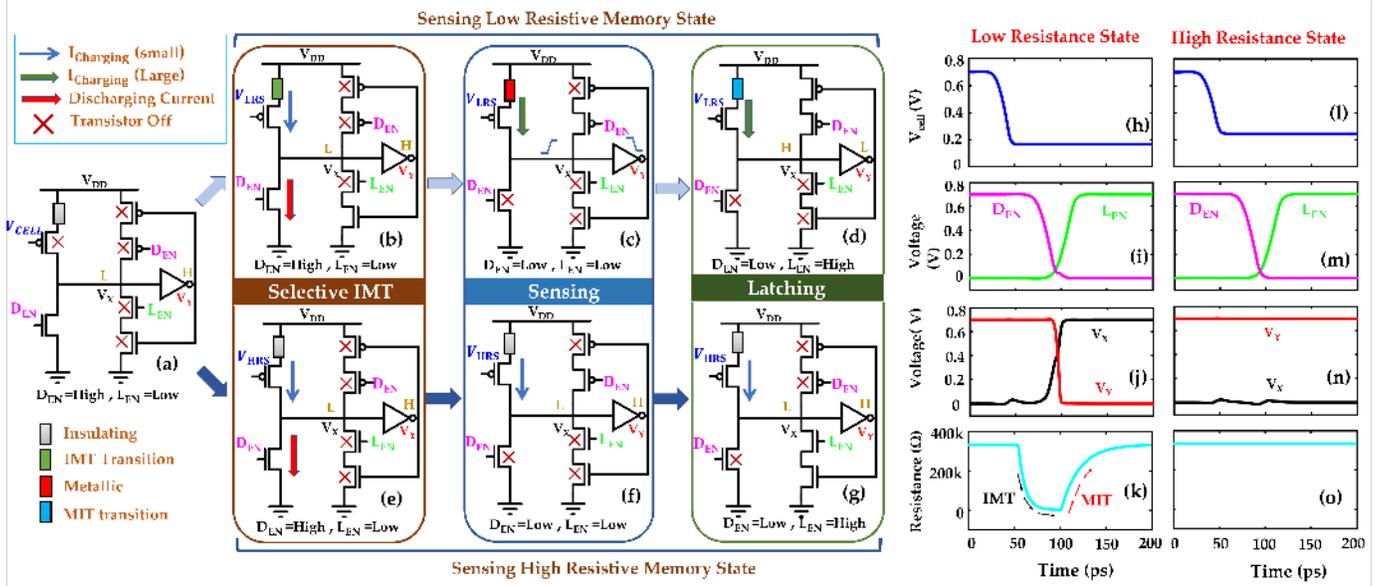

**Figure 7:** Steps of Sensing **(b) -(d)** LRS and **(e)-(g)** HRS for a Hyper-PMOS Voltage Sense Amplifier. Waveforms for sensing **(h)–(k)** LRS and **(l)-(o)** HRS for a Hyper -PMOS VSA.

*A. Hyper-PMOS CSA:*

The schematics of the Hyper-PMOS CSA is illustrated in Fig.5(a). Similar to the traditional CSA, this configuration employs a current-to-voltage converter to generate a voltage corresponding to the current ($I_{CELL}$). In this process, a diode-connected transistor generates a corresponding voltage at the gate node, responding to the input current $I_{CELL}$. Here, for the HRS and LRS of the memory cell, the values of $I_{CELL}$ differ, subsequently leading to two distinct gate-source voltages ($V_{GS\text{-}LRS}$ and $V_{GS\text{-}HRS}$). The host transistor of the Hyper-FET and the PTM is designed in such a way that the relation ($V_{DD} - V_{G\text{-}HRS}$) $< |V_{GS\text{-}IMT}| < (V_{DD} - V_{G\text{-}LRS})$ is satisfied. Thus, when $I_{LRS}$ flows through the transistor, PTM undergoes IMT whereas for $I_{HRS}$, IMT is not induced, thereby keeping the PTM in its insulating state. In this way, the two distinct resistive states of the PTM are leveraged to create two different current levels through the Hyper-FET.

In an instance, where zero cell current is present through the diode-connected PMOS ($I_{CELL} = 0$), the gate-source voltage of both the diode-connected PMOS and the Hyper-PMOS is maintained at 0 ($V_{GS} = 0$). At the beginning of each sensing

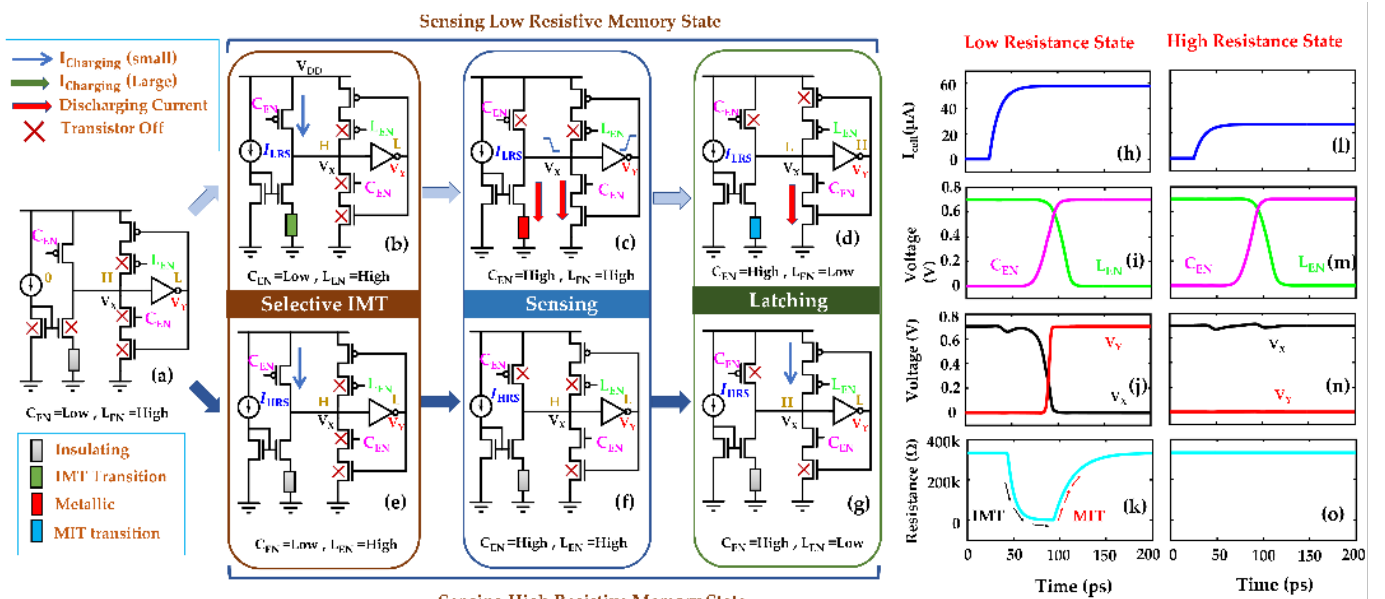

**Figure 8:** Steps of Sensing **(b) -(d)** LRS and **(e)-(g)** HRS for a Hyper-NMOS Current Sense Amplifier. Waveforms for sensing **(h)–(k)** LRS and **(l)-(o)** HRS for a Hyper -NMOS CSA

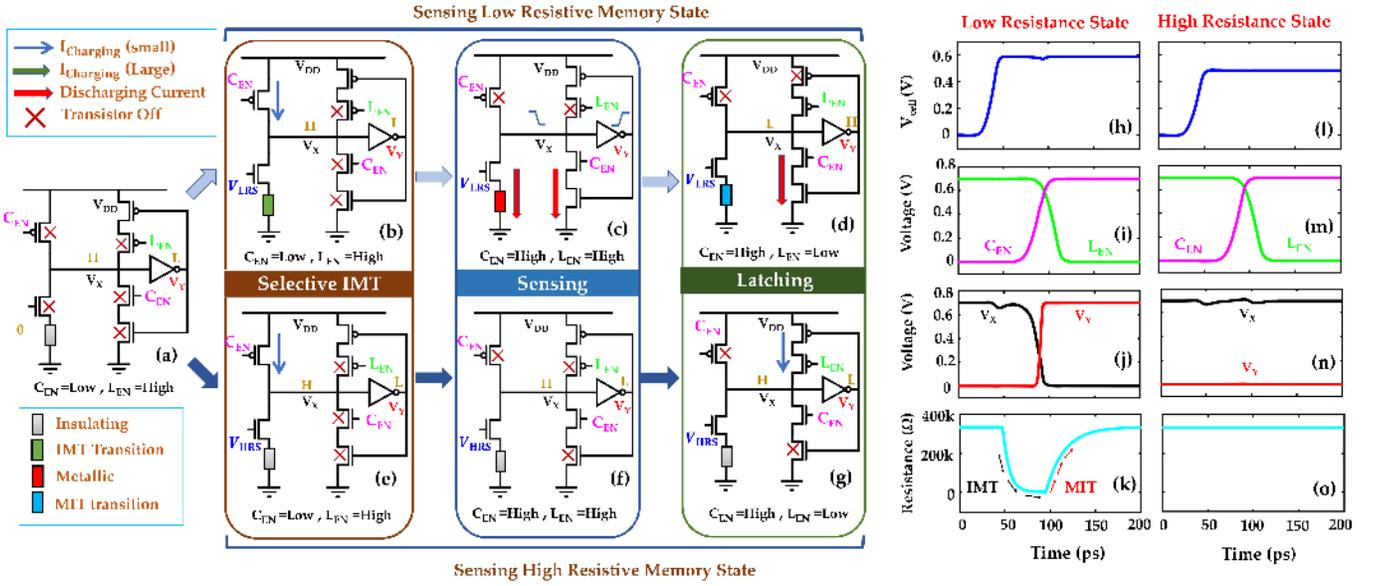

**Figure 9:** Steps of Sensing **(b)**-**(d)** LRS and **(e)**-**(g)** HRS for a Hyper-NMOS Voltage Sense Amplifier. Waveforms for sensing **(h)**–**(k)** LRS and **(l)**-**(o)** HRS for a Hyper -NMOS VSA

cycle, the transistor positioned beneath the Hyper-FET is activated to discharge the 'X' node, setting the stage for the sensing process (Fig.6(b) and Fig.6(e)). So, the voltage at 'X' node ($V_X$) remains grounded (logic '0') and the output node 'Y' remains high (logic '1'). When $I_{CELL}$ flows, the gate voltage of the diode-connected PMOS drops. If the value of the current is significant enough ($\geq I_{LRS}$), the gate to source voltage ($|V_{GS}|$) of the Hyper-FET becomes large enough to trigger IMT (Fig. 6(b)). This prompts a substantial current surge through the Hyper-PMOS. Now, with the de-assertion of the discharge enable ($D_{EN}$) signal, 'X' node begins to be charged up concurrently pulling the output node (node 'Y') through the inverter (Fig.6(c,d)). As $V_Y$ is pulled down, it activates the uppermost PMOS, which accelerates to charge node 'X' (Fig.6(d)). This way, $V_X$ rapidly reaches $V_{DD}$ causing the rapid reduction of the $V_{DS}$ of the Hyper-FET and reducing its drain current. As the current through the PTM reaches $I_{C-MIT}$, MIT occurs and the PTM returns to its metallic state. Consequently, the Hyper-FET ceases to contribute to the charging of node 'X'. Given the low output voltage, the upper PMOS transistor turns on. When $D_{EN}$ is de-asserted, both the 'X' and 'Y' nodes maintain stable voltage values. Figs. 6(h-k) illustrate the simulation results for sensing $I_{LRS}$ in a Hyper-PMOS CSA.

When $I_{HRS}$ flows through the diode connected PMOS, it generates an insufficient gate-source voltage in Hyper-PMOS ($V_{GS} < V_{GS-IMT}$) failing to trigger IMT (Fig. 6(e)). Consequently, the PTM maintains its insulating state resulting in a minimal drain current across the Hyper-FET. Upon the assertion of $D_{EN}$, the pre-charged 'X' node discharges swiftly reducing $V_X$ (Fig. 6(f)). Concurrently, the inverter pulls up $V_Y$. Even if node 'X' experiences a slight increase of charge, the assertion of the latch enable ($L_{EN}$) signal lowers $V_X$ further and $V_Y$ is kept at the high value (Fig. 6(g)). Figs. 6(l-o) illustrate the simulation results for sensing $I_{HRS}$ in a Hyper-PMOS CSA.

### B. Hyper-PMOS VSA:

The schematics of the Hyper-PMOS VSA is illustrated in Fig.5(b). The functionality of the Hyper-PMOS VSA closely resembles that of the Hyper-PMOS CSA except that the current to voltage conversion is not required. Hence, $V_{CELL}$ is directly applied to the gate of the Hyper-PMOS (Fig. 7(a)). Here, the value of the $V_{CELL}$ for the two resistive states are $V_{CELL-LRS}$ and $V_{CELL-HRS}$ respectively directly corresponds to the two different gates to source voltages of the Hyper-PMOS, $V_{GS-LRS}$ and $V_{GS-HRS}$ respectively. Similar to the Hyper-PMOS CSA, meticulous design of the transistor and PTM ensures that the generated gate voltage satisfies the relation, ($V_{DD}$ -$V_{G-HRS}$) < $|V_{GS-IMT}|$ < ($V_{DD}$ - $V_{G-LRS}$). Consequently, for $V_{GS-LRS}$, The PTM undergoes IMT and the metallic state current charges up node X (Figs.7(b,c)). Eventually, node X is charged up to $V_{DD}$ and the inverter pulls down $V_Y$ (Fig. 7(d)). Figs. 7(h-k) illustrate simulation waveform for sensing $V_{LRS}$ in a Hyper-PMOS VSA.

In the case of $V_{HRS}$, IMT does not occur since $|V_{GS}|< |V_{GS-IMT}|$. The small insulating state current of the Hyper-PMOS fails to pull up $V_X$, as upon the de-assertion of $D_{EN}$, the lower NMOS discharges it much faster. Hence, for the $V_{HRS}$, $V_X$ remains low where $V_Y$ remains at the high value (logic '1'). Figs. 7(l-o) illustrate the simulation results of the sensing steps of $V_{HRS}$ for the Hyper-PMOS VSA.

### C. Hyper-NMOS CSA:

The schematic of the Hyper-NMOS CSA is illustrated in Fig.5(b). Here, a diode-connected NMOS is used to mirror the $I_{CELL}$ value to a Hyper-NMOS. Here the host transistor size in the Hyper-NMOS and the PTM is designed in such a way that, $V_{GS-HRS} < V_{GS-IMT} < V_{GS-LRS}$. Similar to the previous topologies, when $I_{LRS}$ flows, the gate voltage of the Hyper-NMOS becomes sufficient to trigger IMT, and a large discharging drain current flow through the Hyper-NMOS (Figs.8(a-d)). This pulls $V_X$ down and with $V_X$ going down, the inverter pulls up the output $V_Y$ (Fig. 8(c)). High $V_Y$ turns on the NMOS at the bottom. When the Charge Enable Signal ($C_{EN}$) is asserted, the NMOS connected with the 'X' node turns on and the two lower NMOSs begin to discharge the 'X' node even faster. As $V_X$ is pulled further, $V_{DS}$ of the Hyper-NMOS goes down reducing the drain

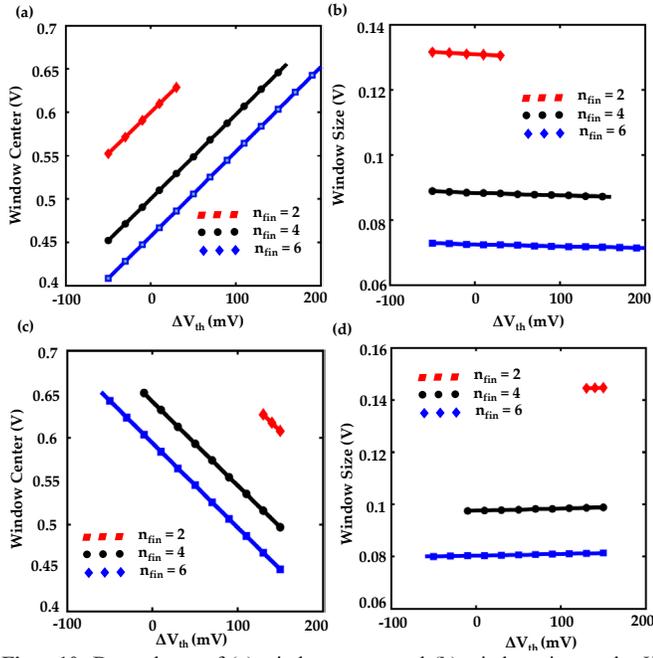

**Fiure 10:** Dependence of **(a)** window center and **(b)** window size on the $V_{th}$ variation for a diode connected NMOS. Dependence of **(c)** window center and **(d)** window size on the $V_{th}$ variation for a diode connected PMOS.

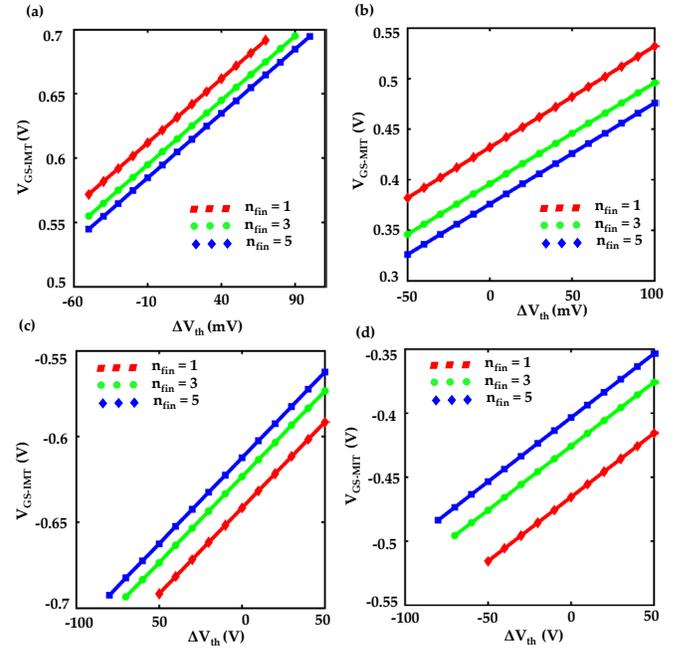

**Figure 11:** Insulator to Metal Transition Voltage, $V_{GS-IMT}$ (V) vs $\Delta V_{th}$ for **(a)** Hyper-NMOS and **(c)** Hyper-PMOS. Metal to Insulator Transition Voltage $V_{GS-MIT}$ (V) vs $\Delta V_{th}$ for **(c)** Hyper-NMOS. **(d)** Hyper-PMOS.

current. Figs. 8(h-k) illustrates the simulation waveform for sensing $I_{LRS}$ for a Hyper-NMOS CSA.

On the other hand, when $I_{CELL}=I_{HRS}$, the generated gate to source voltage of the Hyper-NMOS is $V_{GS-HRS}$. Since, $V_{GS-HRS} < V_{GS-IMT}$, IMT is not triggered and the PTM maintains its insulating state (Fig. 8(e)). Hence, a very small drain current flow through the Hyper-NMOS. With $C_{EN}$ being high at the beginning of the sensing cycle, the small drain current through the Hyper-NMOS fails to discharge node 'X'. When $L_{EN}$ is de-asserted, it creates a charging path from $V_{DD}$ through the upper two PMOS thus keeping node 'X' charged. Consequently, the value of $V_Y$ remains low (Figs. 8(f-g)). Figs. 8(l-o) illustrate the simulation waveform for sensing $I_{HRS}$ for a Hyper-NMOS CSA.

*D. Hyper-NMOS VSA:*

The operation of the Hyper-NMOS VSA closely resembles that of the Hyper-NMOS CSA except that the current to voltage conversion is not required (Fig. 5(d)) and the $V_{CELL}$ is directly applied to the gate of the Hyper-NMOS. Similar to the Hyper-NMOS CSA, the host transistor and PTM are designed in such a way so that the generated gate voltage satisfies the relation, $V_{G-HRS} < V_{GS-IMT} < V_{G-LRS}$. Consequently, for $V_{G-LRS}$, IMT occurs and the metallic state current discharges node X. This pulls down $V_X$ concurrently pulling up $V_Y$ (Figs. 9(a-d)). Figs. 9(h-k) illustrate the the simulation results for sensing $V_{LRS}$ for a Hyper-NMOS VSA.

Conversely, for $V_{CELL} = V_{HRS}$, IMT does not occur, since $V_{GS} < V_{GS-IMT}$. As $C_{EN}$ remains de-asserted at the beginning of sensing, the upper PMOS keeps the 'X' node charged up and the small insulating state current of the Hyper-NMOS fails to pull down $V_X$ (Figs. 9(e-g)). When $L_{EN}$ is de-asserted in the latching cycle, the upper two PMOS keeps $V_X$ high maintaining a low $V_Y$ (Figs. 9(e-g)). Figs. 9(l-o) illustrate the simulation waveforms for sensing $V_{HRS}$ for a Hyper-NMOS VSA.

## V. WINDOW ANALYSIS AND TRANSITION VOLTAGE ANALYSIS

In the proposed CSA topologies, a diode-connected current mirror is used to establish the gate to source voltage ($V_{GS}$) across the Hyper-FET. Hence, it is crucial to analyze how the variation of the mirror transistor affects the generated $V_{GS}$ of the Hyper-FET. The voltages generated by the two levels of $I_{CELL}$ ($I_{LRS}$ and $I_{HRS}$) are $V_{GS-LRS}$ and $V_{GS-HRS}$, respectively. The base transistor and PTM has to be designed in a way that the IMT voltage of the PTM remains between the two values ($|V_{GS-LRS}| < |V_{GS-IMT}| < |V_{GS-HRS}|$)[28].

When the threshold voltage of the transistor ($V_{th}$) is increased, a higher overdrive voltage ($V_{GS}$) is required to generate the same channel current ($I_{CELL}$). So, for the nmos device, both the values ($V_{GS-LRS}$ and $V_{GS-HRS}$) increase for higher $V_{th}$. Thus, the center of the sensing window also increases for the nmos device (Fig. 10(a)) with the increase in $V_{th}$. Similarly, for a pmos device, the overdrive voltage increases ($V_{GS}$ decreases) decreasing the center of the window, as shown in Fig.10(c). Since the source remains at $V_{DD}$, lower $V_{GS}$ is required for higher $V_{th}$. As the drain current through a FinFET increases proportionally with the number of fins [45], the required gate voltage for a particular drain current will be lower for higher number of fins (Figs.10(a-d)). Moreover, the sensitivity of the sensing window to $V_{th}$ is insignificant as shown in Figs. 10(b & d). To create a larger sensing window, the number of fins of the mirror transistor can be reduced (Figs.10(b,d)).

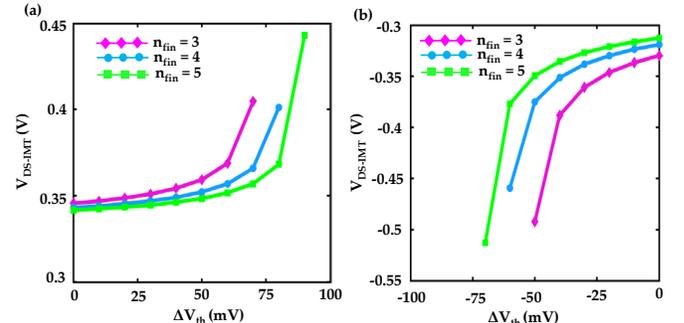

**Figure 12:** Drain to Source Insulator to Metal Transition Voltage $V_{DS-IMT}$ (V) vs threshold voltage variation ($\Delta V_{th}$) for **(a)** Hyper-NMOS **(b)** Hyper-PMOS

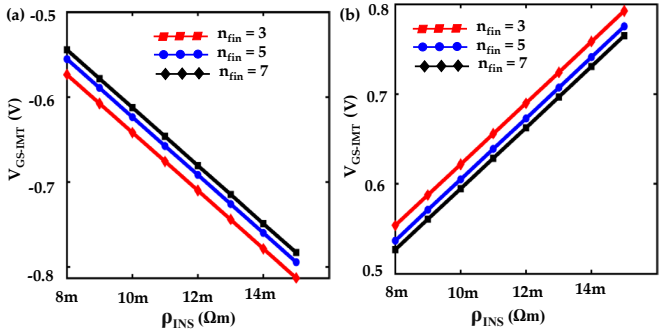

**Figure 13**: Gate-Source Insulator to Metal Transition Voltage $V_{GS-IMT}$ (V) vs Insulator State Resistivity ($\rho_{INS}$) of the PTM for a **(a)** Hyper-NMOS, and **(b)** Hyper-PMOS.

**Table I: Monte-Carlo simulation parameters**

| Parameter | Nominal Value (μ) | Standard Deviation (σ) | Distribution |
|---|---|---|---|
| $R_{INS}$, $R_{MET}$ | 330 kΩ, 6.6 kΩ | 3% | Gaussian (3σ) |
| $V_{C-IMT}$ | 0.336 V | 3% | Gaussian (3σ) |
| $V_{th}$ | 0.2 V | 3.5% | Gaussian (3σ) |

For our proposed topologies, the impact of $V_{GS-IMT}$ is critical as it sets the switching threshold for the sensing operation. With the increase of $V_{th}$, the voltage required for IMT increases as higher $V_{th}$ increases the required overdrive voltage (Figs. 11(a,c)). The dependence of $V_{GS-IMT}$ on $\Delta V_{TH}$ is illustrated in Figs. 11(a,c). The curve shifts rightward for a higher number of fins as the drain current increases for higher number of fins.

The Metal to Insulator transition voltage ($V_{GS-MIT}$) also increases with $V_{th}$ as depicted in Figs. 11(a,c). A higher $V_{th}$ results in a lower metallic state current. Consequently, when we decrease the $V_{GS}$, the PTM current reaches $I_{C-MIT}$ at a higher value of $V_{GS}$. So, with the increase in $V_{th}$, $|V_{GS-MIT}|$ increase for both Hyper-PMOS and Hyper-NMOS (Figs.11(b,d)). In a Hyper-FET, the IMT can be triggered by changing the drain to source voltage as well. Hence, we also examine the dependence of drain voltage switching on $V_{th}$ as shown in Fig.12. We keep $V_{GS}$ at the maximum value ($V_{DD}$) and varied drain voltage to trigger the IMT. As we increase the $V_{DS}$, the drain current and the voltage across the PTM increases. When the voltage across the PTM reaches $V_{C-IMT}$, the PTM undergoes IMT. So, the $V_{DS}$ required for IMT ($V_{DS-IMT}$), increases with the increase in $V_{th}$ for Hyper-NMOS (Fig. 12(a)). At a higher $V_{th}$, the saturation value ($V_{DS-SAT}$) is reached sooner and the increase in $V_{DS}$ doesn't further affect the drain current. Ideally, at a sufficiently high $V_{th}$, IMT fails to occur. However, due to non-ideal effect, saturation current increases with $V_{DS}$ and PTM undergoes IMT at a very high $V_{DS}$ (Fig. 12(a,b)). Finally, we analyze the effect of the insulating state resistivity ($\rho_{INS}$) on the switching voltages. As we increase $\rho_{INS}$, the insulating state current decreases. So, at a higher $\rho_{INS}$, $V_{C-IMT}$ is higher which increases $V_{GS-IMT}$ (Fig.13(a)). Increasing the number of fins increases the current through the PTM. Consequently, for a particular value of $\rho_{INS}$, $V_{GS-IMT}$ is lower for higher number of fins. In the case of the Hyper-PMOS, $V_{GS-IMT}$ and $V_{DS-IMT}$ show similar dependency on $\rho_{INS}$ (Fig. 13(b)).

## VI. VARIATION ANALYSIS

Transistor mismatch poses a significant challenge in conventional SA configurations, particularly in scaled technologies where addressing mismatches within sensing circuits becomes even more challenging. To mitigate the impact of random process variations leading to mismatches, the common approach is to employ larger transistor dimensions. However, this makes the chip size larger and reduces integration density. In our proposed CSA topologies, $I_{CELL}$ is mirrored through P1 and the reference voltage $V_{GS-IMT}$ is generated by P2. So, to ensure lower mismatch, we use 6 fins for P1 and P2. The sensing operation is minimally affected by the variability of the other transistors (N1, N2, N3, P3, P4) as they mostly perform digital operation. Hence, we use 2 fins for these transistors. However, a variation analysis is still required to assess the robustness of our proposed SA topologies.

We perform a 3σ-Monte-Carlo variation analysis to evaluate the sensing delay, power consumption, and power delay product (PDP) for our proposed topologies and compare the results with the conventional SA topologies. We have chosen to vary three significant parameters: (i) IMT voltage ($V_{C-IMT}$), (ii) the length of the PTM ($L_{PTM}$), and (iii) the threshold voltage ($V_{th}$) of P1 and P2. The variation of $L_{PTM}$ affects both the insulating state resistivity ($\rho_{INS}$) and the metallic state resistivity ($\rho_M$). We have incorporated the inverse square root dependency of $V_{th}$ with the number of fins ($n_{fin}$) since transistors with different number of fins were examined in our analysis. While varying one parameter, the others were kept at their nominal values. For conventional SA topologies, we have only included random $V_{th}$ variation. We introduce a 3σ-Gaussian distribution with 5000 data points to the parameters under consideration. The specifications of Monte Carlo Analysis are given in Table I. To measure the time delay for Hyper-PMOS (Hyper-NMOS), we consider the time lag between 50% fall (rise) of $D_{EN}$ ($C_{EN}$) and 50 % fall (rise) of the SA output while sensing LRS.

To estimate the power consumption for a complete sensing cycle, we account the energy dissipation during pre-discharging and latching phases for voltage sensing. Conversely, for the CSAs, we have included the power dissipated in the mirroring the input current in addition to the pre-charging and latching. To compare the overall performance of our proposed SA topologies with conventional SAs, we have calculated power delay product (PDP). The analysis histograms demonstrate that our proposed topologies consistently outperform conventional ones, even in the worst-case scenarios, our designs exhibit better performance in the chosen metrics, as shown in Table II. The worst-case variation is also compared in Table II. Each type of amplifiers (CSA and VSA) is compared with the conventional SAs (Figs. (14-17)). Each type of the SA topologies (CSA and VSA) is compared with the conventional SA topology (Figs. (8-11)). The higher power consumption in the CSAs compared to the VSAs accounts for the extra power required for the current to voltage conversion. Our result indicates that Hyper-PMOS amplifiers exhibit faster sensing, while Hyper-NMOS amplifiers were the most power-efficient option.

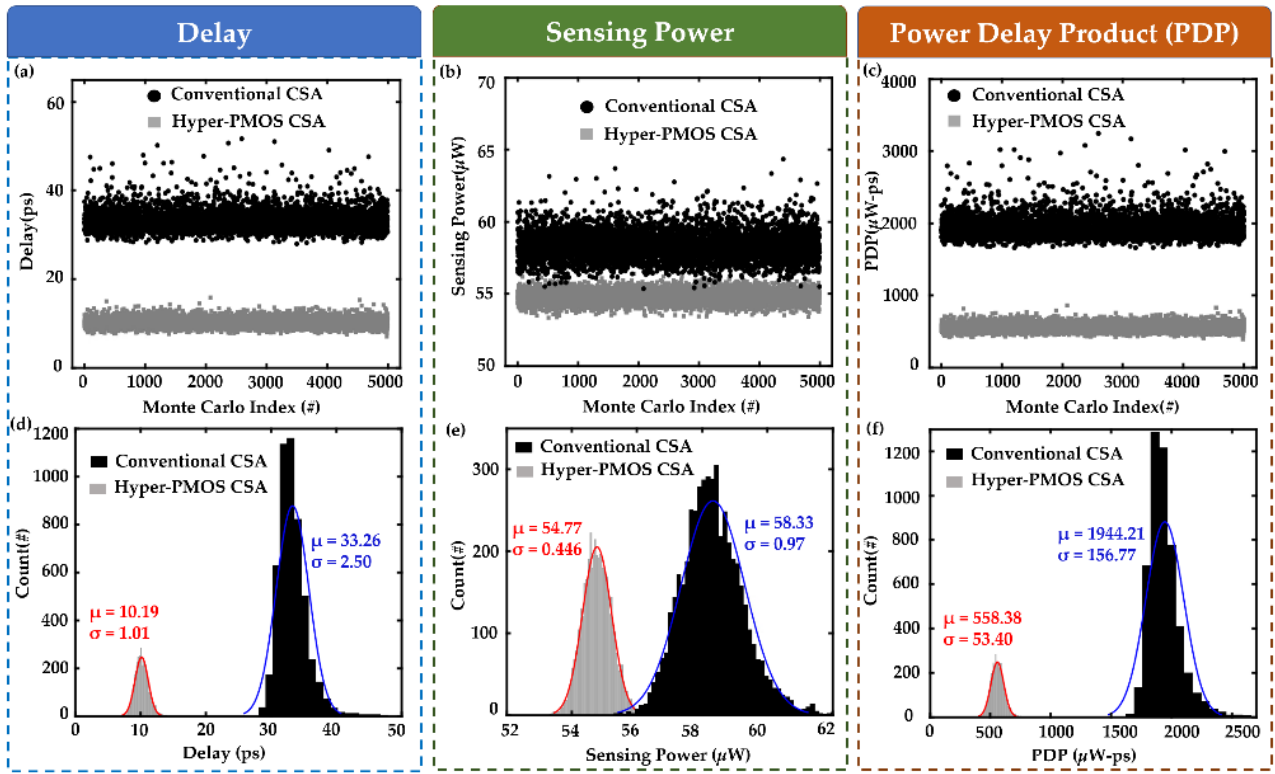

**Figure 14**: Monte Carlo Simulation comparison between Conventional Current Sense Amplifier (CSA) and Hyper-PMOS CSA. Values of **(a)** Delay **(b)** Sensing Power **(c)** Power Delay Product (PDP) and Histogram Comparison of **(d)** Delay **(e)** Sensing Power and **(f)** PDP for different combinations of $V_{th}$, $R_{PTM}$ and $V_{C\text{-}IMT}$

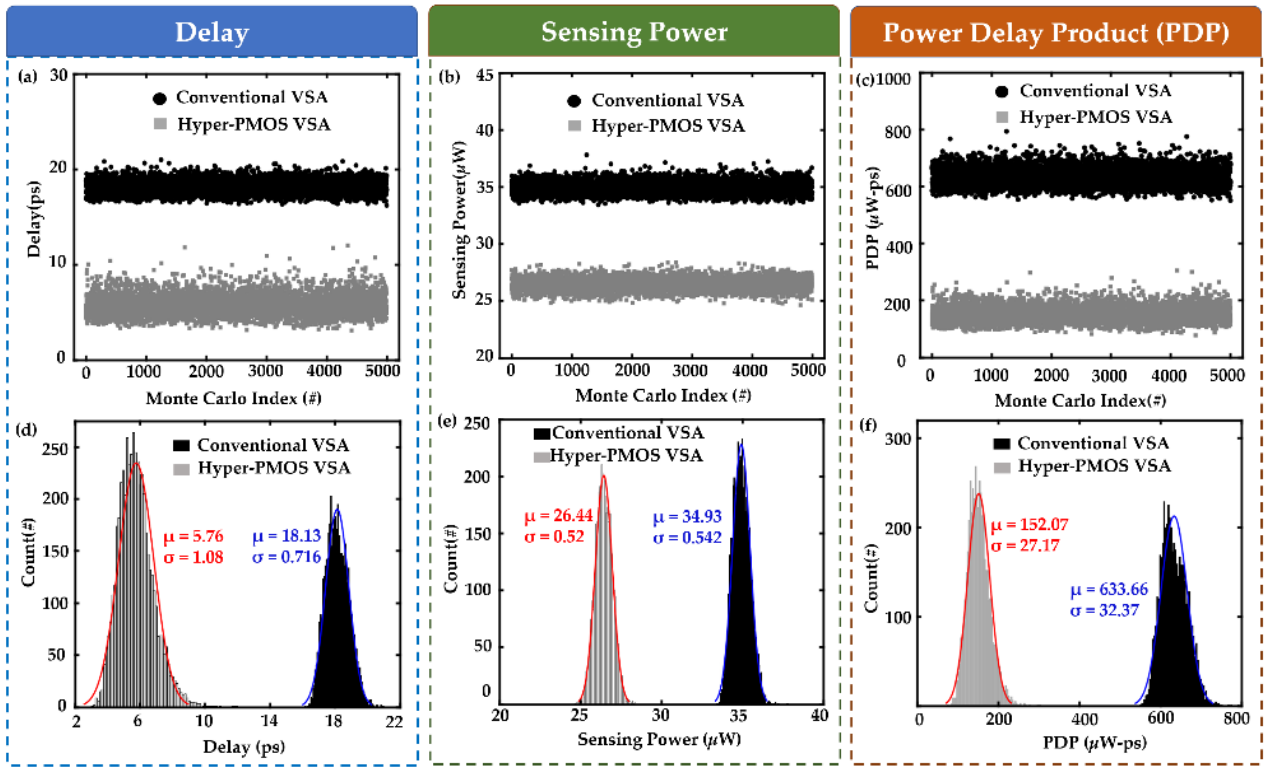

**Figure 15:** Monte Carlo Simulation comparison between Conventional Voltage Sense Amplifier (VSA) and Hyper-PMOS VSA. Values of **(a)** Delay **(b)** Sensing Power **(c)** Power Delay Product (PDP) and Histogram Comparison of **(d)** Delay **(e)** Sensing Power and **(f)** PDP for different combinations of $V_{th}$, $R_{PTM}$ and $V_{C\text{-}IMT}$

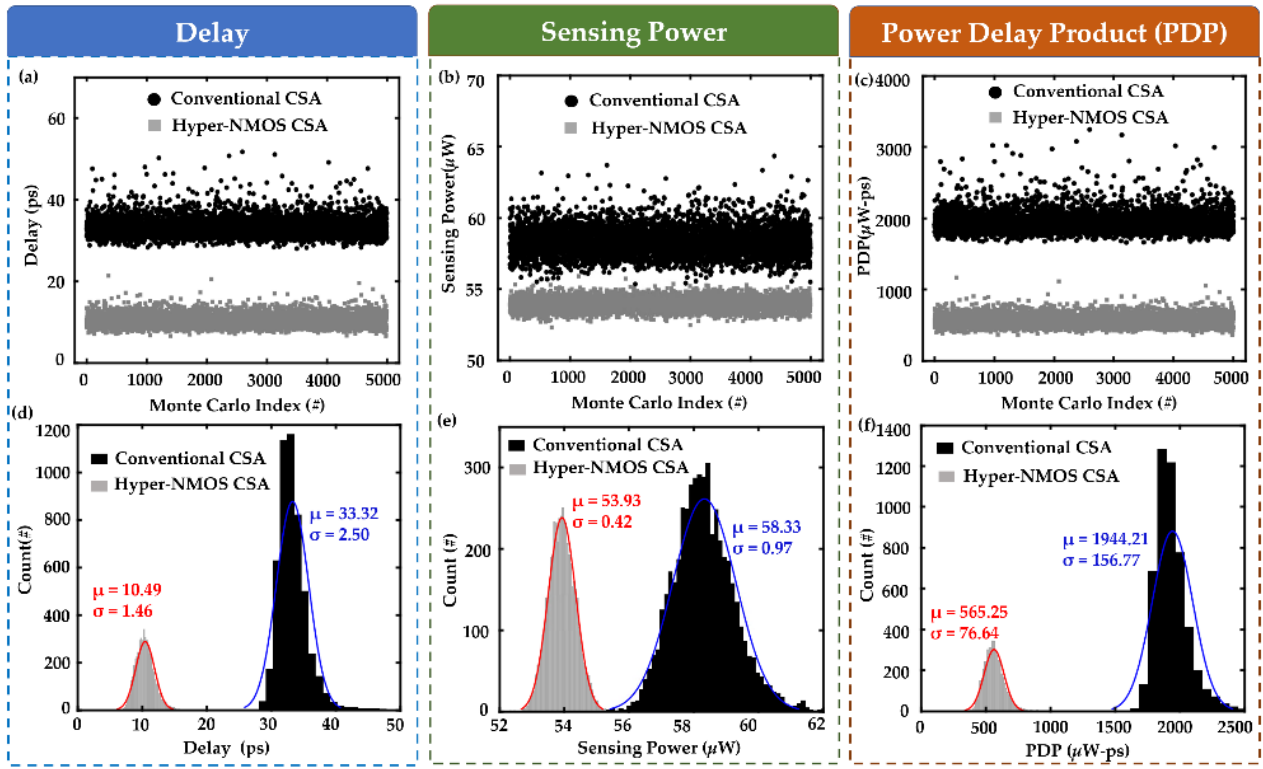

**Figure 16:** Monte Carlo Simulation comparison between Conventional Current Sense Amplifier (CSA) and Hyper-NMOS CSA Values of **(a)** Delay **(b)** Sensing Power **(c)** Power Delay Product (PDP) and Histogram Comparison of **(d)** Delay **(e)** Sensing Power and **(f)** PDP for different combinations of $V_{th}$, $R_{PTM}$ and $V_{C-IMT}$

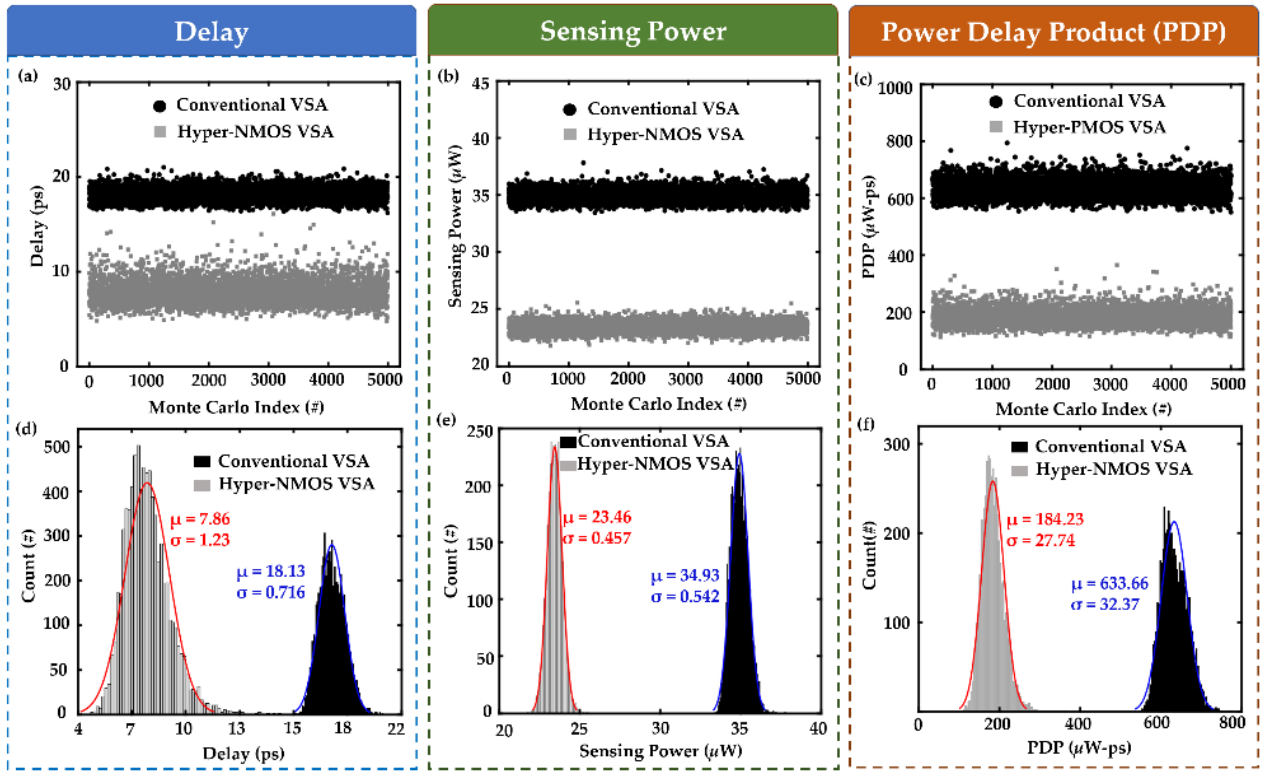

**Figure 17:** Monte Carlo Simulation comparison between Conventional Voltage Sense Amplifier (VSA) and Hyper-PMOS VSA. Values of **(a)** Delay **(b)** Sensing Power **(c)** Power Delay Product (PDP) and Histogram Comparison of **(d)** Delay **(e)** Sensing Power and **(f)** PDP for different combinations of $V_{th}$, $R_{PTM}$ and $V_{C-IMT}$

Table II: Power, Delay and PDP comparison chart for different Sense Amplifier Topology.

| SA Topologies | Figure of Merit | Delay (ps) | | Sensing Power (µW) | | Power Delay Product (PDP) (µW- ps) | |
|---|---|---|---|---|---|---|---|
| | | Average | Worst Case Value | Average | Worst Case Value | Average | Worst Case Value |
| Current Based Sense Amplifiers | Conventional CSA | 33.26 | 104.976 | 58.33 | 64.3368 | 1944.21 | 6539 |
| | Hyper-PMOS CSA | 10.19 | 15.8237 | 54.77 | 56.5764 | 558.33 | 861.4534 |
| | Hyper-NMOS CSA | 10.49 | 21.33 | 53.93 | 55.9065 | 565.25 | 894.554 |
| Voltage Based Sense Amplifiers | Conventional VSA | 18.13 | 20.9865 | 34.93 | 37.8114 | 633.66 | 793.5285 |
| | Hyper-PMOS VSA | 5.76 | 12.013 | 26.44 | 28.3924 | 152.07 | 304.9460 |
| | Hyper-NMOS VSA | 7.86 | 16.09 | 23.46 | 25.56 | 184.23 | 365.735 |

## VII. CONCLUSION

We harnessed the abrupt resistivity change behavior of phase transition material (PTM) to design faster and more efficient sense amplifier circuits capable of sensing both current and voltage in practical scenarios. Since we have chosen the three parameters that are most susceptible to process variation and our variation analysis confirms the enhanced robustness of the proposed SA topologies compared to conventional alternatives. In all of our proposed topologies, any external circuitry to create the reference voltage or current is not required, as it is directly derived from the transition voltages of the PTM. Furthermore, we've showcased a dual means of adjusting the transition voltages, either by changing the host transistor parameters or by merely changing the PTM dimensions. Our developed SA topologies hold significant potential for applications in non-volatile memory, offering promising avenues for achieving high performance and adaptability.

**References**

[1] H. Saito et al., "A chip-stacked memory for on-chip SRAM-rich SoCs and processors," in *IEEE Journal of Solid-State Circuits*, 2010, doi: 10.1109/JSSC.2009.2034078.

[2] N. Chandrasekaran, N. Ramaswamy, and C. Mouli, "Memory technology: Innovations needed for continued technology scaling and enabling advanced computing systems," in *Technical Digest - International Electron Devices Meeting, IEDM*, 2020, doi: 10.1109/IEDM13553.2020.9372125.

[3] S. K. Gupta, S. P. Park, N. N. Mojumder, and K. Roy, "Layout-aware optimization of STT MRAMs," *Proc. -Design, Autom. Test Eur. DATE*, pp. 1455–1458, 2012, doi: 10.1109/date.2012.6176595.

[4] A. Aziz and A. Dissertation, "Device-Circuit Co-Design Employing Phase Transition Materials for Low Power Electronics," Aug. 2019, doi: 10.25394/PGS.8982722.V1.

[5] S. Alam, M. S. Hossain, and A. Aziz, "A cryogenic memory array based on superconducting memristors," *Appl. Phys. Lett.*, vol. 119, no. 8, p. 082602, Aug. 2021, doi: 10.1063/5.0060716.

[6] F. Zahoor, T. Z. Azni Zulkifli, and F. A. Khanday, "Resistive Random Access Memory (RRAM): an Overview of Materials, Switching Mechanism, Performance, Multilevel Cell (mlc) Storage, Modeling, and Applications," *Nanoscale Research Letters*. 2020, doi: 10.1186/s11671-020-03299-9.

[7] M. Le Gallo and A. Sebastian, "An overview of phase-change memory device physics," *Journal of Physics D: Applied Physics*. 2020, doi: 10.1088/1361-6463/ab7794.

[8] S. Alam, M. S. Hossain, S. R. Srinivasa, and A. Aziz, "Cryogenic memory technologies," *Nat. Electron. 2023 63*, vol. 6, no. 3, pp. 185–198, Mar. 2023, doi: 10.1038/s41928-023-00930-2.

[9] S. Alam, M. S. Hossain, and A. Aziz, "A non-volatile cryogenic random-access memory based on the quantum anomalous Hall effect," *Sci. Rep.*, vol. 11, no. 1, pp. 1–9, 2021, doi: 10.1038/s41598-021-87056-7.

[10] S. Alam, M. M. Islam, M. S. Hossain, K. Ni, V. Narayanan, and A. Aziz, "Cryogenic Memory Array based on Ferroelectric SQUID and Heater Cryotron," *2022 Device Res. Conf.*, pp. 1–2, Jun. 2022, doi: 10.1109/DRC55272.2022.9855813.

[11] A. Grossi et al., "Resistive RAM endurance: Array-level characterization and correction techniques targeting deep learning applications," *IEEE Trans. Electron Devices*, 2019, doi: 10.1109/TED.2019.2894387.

[12] A. Zeumault, S. Alam, Z. Wood, R. J. Weiss, A. Aziz, and G. S. Rose, "TCAD Modeling of Resistive-Switching of HfO2 Memristors: Efficient Device-Circuit Co-Design for Neuromorphic Systems," *Front. Nanotechnol.*, vol. 3, p. 71, 2021, doi: 10.3389/FNANO.2021.734121.

[13] A. Aziz and S. K. Gupta, "Threshold Switch Augmented STT MRAM: Design Space Analysis and Device-Circuit Co-Design," *IEEE Trans. Electron Devices*, vol. 65, no. 12, pp. 5381–5389, Dec. 2018, doi: 10.1109/TED.2018.2873738.

[14] S. George et al., "Nonvolatile memory design based on ferroelectric FETs," *Proc. - Des. Autom. Conf.*, vol. 05-09-June-2016, Jun. 2016, doi: 10.1145/2897937.2898050.

[15] A. Zeumault, S. Alam, M. O. Faruk, and A. Aziz, "Memristor compact model with oxygen vacancy concentrations as state variables," *J. Appl. Phys.*, vol. 131, no. 12, p. 124502, Mar. 2022, doi: 10.1063/5.0087038.

[16] J. Hutchins et al., "A Generalized Workflow for Creating Machine Learning-Powered Compact Models for Multi-State Devices," *IEEE Access*, vol. 10, pp. 115513–115519, 2022, doi: 10.1109/ACCESS.2022.3218333.

[17] S. T. Han, Y. Zhou, and V. A. L. Roy, "Towards the development of flexible non-volatile memories," *Advanced Materials*. 2013, doi: 10.1002/adma.201301361.

[18] Divya and P. Mittal, "A low-power high-performance voltage sense amplifier for static RAM and comparison with existing current/voltage sense amplifiers," *Int. J. Inf. Technol.*, 2022, doi: 10.1007/s41870-022-00916-x.

[19] H. L. Chee, Y. Z. Kok, T. N. Kumar, and H. A. F. Almurib, "Sense amplifier for ReRAM-based crossbar memory systems," *Int. J. Electron. Lett.*, 2023, doi: 10.1080/21681724.2022.2067903.

[20] S. Alam, M. M. Islam, M. S. Hossain, and A. Aziz, "Superconducting Josephson Junction FET-based Cryogenic Voltage Sense Amplifier," *2022 Device Res. Conf.*, pp. 1–2, Jun. 2022, doi: 10.1109/DRC55272.2022.9855654.

[21] L. Zhang et al., "A novel sense amplifier to mitigate the impact of NBTI and PVT variations for STT-MRAM," *IEICE Electron. Express*, 2019, doi: 10.1587/elex.16.20190238.

[22] M. Li, J. F. Kang, and Y. Y. Wang, "A novel voltage-type sense amplifier for low-power nonvolatile memories," *Sci. China, Ser. F Inf. Sci.*, 2010, doi: 10.1007/s11432-010-4015-8.

[23] S. Motaman, S. Ghosh, and J. P. Kulkarni, "A novel slope detection technique for robust STTRAM sensing," *Proc. Int. Symp. Low Power Electron. Des.*, vol. 2015-Septe, pp. 7–12, 2015, doi: 10.1109/ISLPED.2015.7273482.

[24] M. F. Chang et al., "Challenges and circuit techniques for energy-efficient on-chip nonvolatile memory using memristive devices," *IEEE J. Emerg. Sel. Top. Circuits Syst.*, vol. 5, no. 2, pp. 183–193, Jun. 2015, doi: 10.1109/JETCAS.2015.2426531.

[25] M. F. Chang *et al.*, "An offset-tolerant fast-random-read current-sampling-based sense amplifier for small-cell-current nonvolatile memory," *IEEE J. Solid-State Circuits*, vol. 48, no. 3, pp. 864–877, 2013, doi: 10.1109/JSSC.2012.2235013.
[26] S. Alam *et al.*, "Threshold Switch Assisted Memristive Memory with Enhanced Read Distinguishability," pp. 531–534, Nov. 2022, doi: 10.1109/NANO54668.2022.9928710.
[27] N. Shukla *et al.*, "A steep-slope transistor based on abrupt electronic phase transition," *Nat. Commun.*, vol. 6, no. 1, pp. 1–6, Aug. 2015, doi: 10.1038/ncomms8812.
[28] A. Aziz *et al.*, "Low power current sense amplifier based on phase transition material," *Device Res. Conf. - Conf. Dig. DRC*, vol. 63, no. 12, pp. 2016–2017, 2017, doi: 10.1109/DRC.2017.7999425.
[29] A. Aziz, N. Shukla, S. Datta, and S. K. Gupta, "COAST: Correlated material assisted STT MRAMs for optimized read operation," *Proc. Int. Symp. Low Power Electron. Des.*, vol. 2015-September, pp. 1–6, Sep. 2015, doi: 10.1109/ISLPED.2015.7273481.
[30] J. Vaidya, R. S. S. Kanthi, S. Alam, N. Amin, A. Aziz, and N. Shukla, "A Three-terminal Non-Volatile Ferroelectric Switch with an Insulator-Metal Transition Channel," Aug. 2021.
[31] S. Alam *et al.*, "Design Space Exploration for Threshold Switch Assisted Memristive Memory," *IEEE Trans. Nanotechnol.*, pp. 1–8, 2023, doi: 10.1109/TNANO.2023.3305931.
[32] N. Shukla *et al.*, "Ag/HfO2 based threshold switch with extreme non-linearity for unipolar cross-point memory and steep-slope phase-FETs," *Tech. Dig. - Int. Electron Devices Meet. IEDM*, pp. 34.6.1-34.6.4, 2017, doi: 10.1109/IEDM.2016.7838542.
[33] L. D. R. Lq *et al.*, "2II FXUUHQW aS $ DQG + LJK ( QGXUDQFH !," pp. 253–256, 2015.
[34] S. V. Streltsov and D. I. Khomskii, "Orbital-dependent singlet dimers and orbital-selective Peierls transitions in transition-metal compounds," *Phys. Rev. B - Condens. Matter Mater. Phys.*, 2014, doi: 10.1103/PhysRevB.89.161112.
[35] R. Sakuma, T. Miyake, and F. Aryasetiawan, "First-principles study of correlation effects in VO2," *Phys. Rev. B - Condens. Matter Mater. Phys.*, 2008, doi: 10.1103/PhysRevB.78.075106.
[36] A. Pashkin *et al.*, "Ultrafast insulator-metal phase transition in VO2 studied by multiterahertz spectroscopy," *Phys. Rev. B - Condens. Matter Mater. Phys.*, 2011, doi: 10.1103/PhysRevB.83.195120.
[37] S. Alam, M. M. Islam, A. Jaiswal, N. Cady, G. Rose, and A. Aziz, "Variation-aware Design Space Exploration of Mott Memristor-based Neuristors," *Proc. IEEE Comput. Soc. Annu. Symp. VLSI, ISVLSI*, vol. 2022-July, pp. 68–73, 2022, doi: 10.1109/ISVLSI54635.2022.00025.
[38] M. M. Islam, M. Hernandez Rivero, G. Rose, and A. Aziz, "Low-Power Dynamic Circuit Design with Steep-Switching Hybrid Phase Transition FETs (Hyper-FETs)," *IEEE Trans. Electron Devices*, vol. 70, no. 2, pp. 819–825, Feb. 2023, doi: 10.1109/TED.2022.3231236.
[39] A. Aziz, N. Shukla, S. Datta, and S. K. Gupta, "Steep switching hybrid phase transition FETs (Hyper-FET) for low power applications: A device-circuit co-design perspective - Part II," *IEEE Trans. Electron Devices*, vol. 64, no. 3, pp. 1358–1365, 2017, doi: 10.1109/TED.2017.2650598.
[40] S. Alam, W. M. Hunter, N. Amin, M. M. Islam, S. K. Gupta, and A. Aziz, "Design Space Exploration for Phase Transition Material Augmented MRAMs With Separate Read-Write Paths," *IEEE Trans. Comput. Des. Integr. Circuits Syst.*, 2023, doi: 10.1109/TCAD.2023.3299838.
[41] X. Peng, R. Madler, P. Y. Chen, and S. Yu, "Cross-point memory design challenges and survey of selector device characteristics," *J. Comput. Electron.*, 2017, doi: 10.1007/s10825-017-1062-z.
[42] R. A. Cernea *et al.*, "A 34 MB/s MLC write throughput 16 Gb NAND with all bit line architecture on 56 nm technology," in *IEEE Journal of Solid-State Circuits*, 2009, vol. 44, no. 1, pp. 186–194, doi: 10.1109/JSSC.2008.2007152.
[43] M. F. Chang, S. M. Yang, C. W. Liang, C. C. Chiang, P. F. Chiu, and K. F. Lin, "Noise-immune embedded NAND-ROM using a dynamic split source-line scheme for VDDmin and speed improvements," in *IEEE Journal of Solid-State Circuits*, 2010, vol. 45, no. 10, pp. 2142–2155, doi: 10.1109/JSSC.2010.2060279.
[44] J. Zhu, N. Bai, and J. Wu, "Review of sense amplifiers for static random access memory," *IETE Technical Review (Institution of Electronics and Telecommunication Engineers, India)*, vol. 30, no. 1. pp. 72–81, Jan-2013, doi: 10.4103/0256-4602.107343.
[45] M. Jurczak, N. Collaert, A. Veloso, T. Hoffmann, and S. Biesemans, "Review of FINFET technology," in *Proceedings - IEEE International SOI Conference*, 2009, doi: 10.1109/SOI.2009.5318794.